\documentclass[preprint,authoryear,12pt]{elsarticle}

\usepackage{amssymb}

\usepackage{rotating}

\journal{Planetary and Space Science}

\newcommand{\tento}[1]{\cdot 10^{#1}}

\newcommand{\Dp}{$\Delta_{P_{\rm vap}}$}
\newcommand{\Dv}{$\Delta_{V_{\rm m}}$}
\newcommand{\Dh}{$\Delta_{\Delta H_{\rm vap}}$}
\newcommand{\Dl}{$\Delta_{l_{ij}}$}
\newcommand{\DA}{$\Delta_{\rm All}$}

\newcommand{\Sp}{$\sigma^{*}_{\small P_{\rm vap}}$}
\newcommand{\Sv}{$\sigma^{*}_{\small V_{\rm m}}$}
\newcommand{\Sh}{$\sigma^{*}_{\small \Delta H_{\rm vap}}$}
\newcommand{\Sl}{$\sigma^{*}_{\small l_{ij}}$}
\newcommand{\SA}{$\sigma^{*}_{\small \rm All}$}

\usepackage{color}
\definecolor{backcolor}{rgb}{0.89,0.58,0.41}

\begin{document}

\begin{frontmatter}



\title{Titan's lakes chemical composition: sources of uncertainties and variability}


\author[ipr,enscr,ueb]{D.~Cordier}
\ead{daniel.cordier@ensc-rennes.fr}

\author[UTINAM]{O.~Mousis}
\author[URome,LPL]{J. I.~Lunine}
\author[LMD]{S.~Lebonnois}
\author[REIMS]{P.~Rannou}
\author[LPL]{P.~Lavvas}
\author[COIMBRA]{L.Q.~Lobo}
\author[COIMBRA]{A.G.M.~Ferreira}

\address[ipr]{Institut de Physique de Rennes, CNRS, UMR 6251, Universit{\'e} de Rennes 1, Campus de Beaulieu, 35042 Rennes, France}

\address[enscr]{Ecole Nationale Sup{\'e}rieure de Chimie de Rennes, CNRS, UMR 6226, Avenue du G\'en\' eral Leclerc,
                CS 50837, 35708 Rennes Cedex 7, France}

\address[ueb]{Universit\'e europ\'eenne de Bretagne, Rennes, France}

\address[UTINAM]{Universit{\'e} de Franche-Comt{\'e}, Institut UTINAM, CNRS/INSU, UMR 6213, 25030 Besan\c{c}on Cedex, France}

\address[URome]{Dipartimento di Fisica, Universit{\`a} degli Studi di Roma ``Tor Vergata'', Rome, Italy}

\address[LPL]{Lunar and Planetary Laboratory, University of Arizona, Tucson, AZ, USA}

\address[LMD]{Laboratoire de M\'et\'eorologie Dynamique, Jussieu, Box 99, 75252 PARIS cedex 05, France}

\address[REIMS]{Groupe de Spectrom\'etrie Mol\'eculaire et Atmosph\'erique - UMR 6089 
Campus Moulin de la Housse - BP 1039
Universit\'e de Reims Champagne-Ardenne
51687 REIMS -- France}

\address[COIMBRA]{Departamento de Engenharia Quimica, Universidade de Coimbra, Coimbra 3030-290, Portugal}

\begin{abstract}
   Between 2004 and 2007 the instruments of the CASSINI spacecraft, orbiting within the  Saturn system, discovered dark patches in the polar regions of Titan. These features are interpreted as hydrocarbon lakes and seas with ethane and methane identified as the main compounds. In this context, we have developed a lake-atmosphere equilibrium model allowing the determination of the chemical composition of these liquid areas present on Titan. The model is based on uncertain thermodynamic data and precipitation rates of organic species predicted to be present in the lakes and seas that are subject to spatial and temporal variations. Here we explore and discuss the influence of these uncertainties and variations. The errors and uncertainties relevant to thermodynamic data are simulated via Monte-Carlo simulations. Global Circulation Models (GCM) are also employed
in order to investigate the possibility of chemical asymmetry between the south and the north poles, due to differences in precipitation rates. 
 We find that mole fractions of compounds in the liquid phase have a high sensitivity to thermodynamic data used as inputs, in particular molar volumes and enthalpies of vaporization. When we combine all considered uncertainties, the ranges of obtained mole fractions are rather large (up to $\sim$ 8500\%) but the distributions of values are narrow. The relative standard deviations remain between 10\% and $\sim 300$\% depending on the compound 
considered. Compared to other sources of uncertainties and variability, deviation caused by surface pressure variations are clearly negligible, remaining of the order of a few percent up to $\sim20$\%. Moreover no significant difference is found between the composition of lakes located in north and south poles.
Because the theory of regular solutions employed here is sensitive to thermodynamic data and is not suitable for polar molecules such as HCN and CH$_{3}$CN, our work strongly underlines the need for experimental simulations and the improvement of Titan's atmospheric models. 
\end{abstract}


\end{frontmatter}


\section{Introduction}
\label{sec:intro}

 The surface of Saturn's haze-shrouded moon Titan had long been proposed to be at least partly hidden by oceans or seas, on the basis of the stability of liquid methane and ethane at the surface  \citep[][]{flasar_1983,lunine_etal_1983,lorenz_etal_2003}. The presence of a global ocean on Titan was excluded from ground-based radar observations in the mid 1990s \citep{muhleman_etal_1995}. In mid 2006,  dark, lake-like features of a range of sizes were detected at Titan's north polar region by the Cassini RADAR \citep{stofan_etal_2007}. The chemical composition of these lakes remains, however, poorly determined. Spectra of the southern hemisphere lake Ontario Lacus have been obtained by the Visual and Infrared Mapping Spectrometer (VIMS) aboard Cassini but the only species that has been firmly identified is C$_2$H$_6$ \citep{brown_etal_2008}. The difficulty in determining the composition of the lakes is essentially due to the presence of a large atmospheric fraction of CH$_4$ that impedes this molecule's identification in the liquid phase present on the surface, irrespective of the value of its mole fraction. However, methane is indirectly inferred in Ontario Lacus by the secular decline of the lake extent over the Titan summer \citep{hayes_etal_2010} and the observation of troposphere clouds, which must be methane, coincident with surface darkening over the southern pole during the summer \citep{turtle_etal_2007}. Because the detection of other compounds in the lakes of Titan remains challenging in the absence of in situ measurements, the only way to get a good estimate of their chemical composition is to develop and utilize a thermodynamic model based on theoretical calculations and laboratory data. Several models investigating the influence of photochemistry and the atmospheric composition on the chemical composition of putative hydrocarbon oceans or seas formed on the surface of Titan, have been elaborated in the pre-Cassini years 
\citep[][]{lunine_etal_1983,dubouloz_etal_1989,mckay_etal_1993,tokano_2005}. These models suggested that the liquid phase existing on Titan contains a mixture made from C$_2$H$_6$, CH$_4$ and N$_2$, and a large number of dissolved minor species.\\

On the other hand, the Cassini-Huygens measurements have improved our knowledge of the structure and composition of Titan's atmosphere. In particular, the Gas Chromatograph Mass Spectrometer (GCMS) aboard Huygens and the Cassini Composite Infrared Spectrometer (CIRS) provided new atmospheric mole fraction data \citep[see][and Table 1]{niemann_etal_2005}.
Moreover, near-surface brightness temperatures and corresponding estimates for physical temperatures in the high latitudes at which numerous lakes are found have now been determined \citep{jennings_etal_2009}. These atmospheric and surface conditions have been recently used to recompute the solubilities of the different compounds in the hydrocarbon lakes \citep[][hereafter C09]{cordier_etal_2009}. The same model has also been employed to explore the possibility of noble gas trapping in the lakes of Titan in order to provide an attempt of explanation of their atmospheric depletion \citep{cordier_etal_2010}. 

The assumptions considered by C09 are similar to those made by \cite{dubouloz_etal_1989} (hereafter DUB89): in both cases, lakes are considered as nonideal solutions in thermodynamic equilibrium with the atmosphere. However, neither DUB89 nor C09 have taken into consideration the influence of uncertainties on the data used as inputs in their models. Indeed, some thermodynamic data are measured at much higher temperature and extrapolated down to 
temperatures relevant to Titan's conditions. Precipitation rates are also supposed to vary with respect to latitude, longitude and time. 
In this work, we investigate the influence of thermodynamic uncertainties, and in a lesser extent, the geographic influence of the variation of precipitations
on the lakes composition. In the latter case, we restrict our study to a supposed north/south poles
asymmetry in chemical composition.

      In Section~\ref{model} we detail our lake-atmosphere equilibrium model. Section~\ref{thermo} is dedicated to the study of the influence of uncertainties on thermodynamic data
(vapor pressures, molar volumes, enthalpies of vaporization and parameters of interaction)  on the resulting lakes composition. 
In Sect.~\ref{precip}, simulations are conducted with the use of precipitation rates derived from a version of the
IPSL\footnote{Institut Pierre-Simon Laplace} 2-dimensional climate model of Titan's atmosphere
\citep{crespin_etal_2008} and allow comparison between chemical composition of south pole and north pole lakes.
Section~\ref{discuss} is devoted to discussion and conclusions. 

\section{\label{model}Description of the lake-atmosphere equilibrium model}

Our model is based on regular solution theory and thermodynamic equilibrium
is assumed between the liquid and the atmosphere. This equilibrium, which is expressed by the equality of chemical potentials,
can be written as follows (Eq. 1 of DUB89):

\begin{equation}\label{equa1}
 Y_{k} \, P = \Gamma_{k} \, X_{k} \, P_{vp,k},
\end{equation}

\noindent where $P$ is the total pressure at Titan's surface, $Y_{k}$ and $X_{k}$ respectively the mole fractions of the $k$ compound in the atmosphere and in the liquid, and $P_{vp,k}$ its vapor pressure. The activity coefficient $\Gamma_{k}$ (dimensionless) of the $k$ compound is given by (frame of the regular solution theory -- see  \cite{poling_2007}):

\begin{equation}\label{coeff_activity}
R T \, \mathrm{ln} \, \Gamma_{k} = V_{m,k} \sum_{i} \sum_{j} (A_{ik}-A_{ij}) \, \Phi_{i} \Phi_{j}
\end{equation}

where

\begin{equation}
 A_{ij} = (\delta_{i}-\delta_{j})^{2} + 2 l_{ij} \, \delta_{i} \delta_{j}
\end{equation}

and

\begin{equation}
 \label{equaphii}
 \Phi_{i} = X_{i} V_{m, i} / \sum_{j} X_{j} \, V_{m, j}.
\end{equation}

\noindent $\delta_{i}$ ($(\mathrm{J.m}^{-3})^{1/2}$) 
is the Hildebrand's solubility parameter of the $i$th compound. The value of this parameter is given by:

\begin{equation}
 \delta_{i} = \sqrt{\frac{\displaystyle\Delta H_{v,i} - R T}{\displaystyle V_{m, i}}}
\end{equation}

\noindent where $\Delta H_{v,i}$ ($\mathrm{J\,mol}^{-1}$) is the enthalpy of vaporization and $V_{m,i}$ ($\mathrm{m^{3}\,mol^{-1}}$) the molar volume.
A $\delta_{i}$ represents a measure of the molecular cohesion energy of the pure component $i$. It depends on the nature and the
{\bf strength} of intermolecular forces (hydrogen bond, ...) between molecules of the same species. In general, two components $i$ and $j$ with
$\delta_{i}$ and $\delta_{j}$ presenting close values, have a high solubility. Beside this,
the $l_{ij}$'s parameters represent the effects of interactions between molecules of different species. These
$l_{ij}$'s are empirically determined and are generally poorly known. The situation $\forall i, j$: $\delta_{i} = \delta_{j}$
and $l_{ij}= 0$ corresponds to all activity coefficient equal to one, in other words this is an ideal solution in which all
intermolecular forces are negligible.

Our model also allows us to estimate the mole fraction of each solid precipitate that is dissolved in the lakes of Titan. To this end, we calculate the \textit{saturation} mole fraction\footnote{The saturation mole fraction of the compound $i$ corresponds to the maximum mole fraction of $i$ in the liquid form. Above this value, the $i$ material in excess remains in solid form.}
$X_{i,sat}$ of the compound $i$, which is given by (Eq. 7 of DUB89):

\begin{equation}\label{equa2}
\mathrm{ln}(\Gamma_{i} \,  X_{i,sat}) = (\Delta H_{m}/R T_{m})(1-T_{m}/T),
\end{equation}

\noindent where $T_m$ is the component's melting temperature and $\Delta H_m$ its enthalpy of fusion. Our calculation procedure is then {\bf} as follows: 

\begin{enumerate}
   \item The unknown $X_{i}$'s and $Y_{i}$'s are computed via the Newton-Raphson method.
   \item Once the $X_{i}$'s have been determined, the $X_{i,sat}$'s are in turn calculated and compared to the $X_{i}$'s for each species. 
          If for $i$ compound we get $X_{i,sat}~<~X_{i}$, then we fix $X_{i} = X_{i,sat}$.
   \item We get new values of $X_{i}$'s and $X_{i,sat}$'s via the resolution of the nonlinear system.
   \item The iterations are continued until we get a difference between $X_{i,sat}$ and $X_{i}$ lower than $10^{-6}$, value for which the numerical
         inaccuracy is clearly negligible compared to other sources of uncertainties.
\end{enumerate}

\noindent The known $Y_{i}$'s are given in Table~\ref{atmoscompo}. 
 The precipitation rates $\tau_{i}$'s represent the number of molecules of a given species, reaching the surface of Titan by unit of time and by unit
of surface (molecules \, m$^{-2}$ \, s$^{-1}$).
The  $\tau_{i}$'s used in C09 were derived from the photochemical models of \cite{lavvas_etal_2008a,lavvas_etal_2008b} and \cite{vuitton_etal_2008}, and correspond to the main products of CH$_4$ and N$_2$ photolysis. 
These rates allow us to express each $i$ compound that falls from the atmosphere in the form 

\begin{equation}
\label{eqprecipit}
X_{i} = \frac{\tau_i}{\tau_{C_2H_6}} \times X_{C_2H_6}
\end{equation}

\noindent We also ensure that $\sum_{i} X_i$ = 1 and $\sum_{i} Y_i$ = 1. In this way, we get 15 unknowns and 15 equations, allowing the system to be solved. The thermodynamic data used in our calculations derive mainly from the NIST database\footnote{\texttt{http://webbook.nist.gov}}. As discussed in the following section, these data are often not well known and the large uncertainties associated to their determination may induce strong variations of the lakes chemical composition. 

\section{\label{thermo}Uncertainties due to thermodynamic data}

In the calculations of C09, the thermodynamic data (vapor pressures, molar volumes, enthalpies of vaporization and $l_{ij}$'s) have been set to their nominal values in the lake-atmosphere equilibrium model. However, each of these nominal values is accompanied by a given ``deviation'' or ``error'' and the consideration of the full range of possibilities for these thermodynamic data may strongly alter the lakes composition compared to the one calculated by C09. In order to investigate up to which point the composition of these lakes may depart from the one of C09, we use here a Monte-Carlo numerical method \citep[see for instance][]{metropolis_ulam_1949}, allowing us to perform error simulations for vapor pressures, molar volumes, enthalpies of vaporization and parameters of interaction $l_{ij}$'s. A first set of computations consists in calculating the composition of lakes from numbers randomly chosen within the range of possible values attributed to a set of thermodynamic data. For each mole fraction $X_{i}$, the minimum $X_{i, min}$, the maximum $X_{i,max}$ and the average value $\overline{X}_{i}$  are recorded. The procedure is repeated 10,000 times for each set of thermodynamic data. The choice of the total number of Monte-Carlo iterations is a compromise that has been fixed to get a statistically significant population while maintaining reasonable computation time. Simulations with 5,000 and 20,000 iterations do not give significantly different results. Additionally we perform a simulation addressing the case where all sets of thermodynamic data are simultaneously considered with synthetic errors. This procedure is also repeated 10,000 times.

Note that the enthalpies of melting $\Delta H_{\rm m}$ are not considered in our investigation because i) they seem to be reasonably well known compared to other thermodynamic quantities and ii) they only play a role in case of saturation, e.g here with HCN. For this compound the measurement, provided by the NIST database, comes from \cite{giauque_ruehrwein_1939} with an accuracy of about $10^{-4}$. Such a level of uncertainty is clearly negligible compared to other sources, allowing us to keep these thermodynamic quantities out of our study.

\subsection{\label{Pvap}Influence of vapor pressure uncertainties}

Vapor pressures of species, for which Eq.~\ref{equa1} is written, are taken from the NIST database in the form of an Antoine's law in the cases of N$_{2}$, CH$_{4}$, Ar and C$_{2}$H$_{6}$ or from a vapor pressure law given by \cite{handbook74th} in the case of CO\footnote{CO data are  unavailable in the NIST.}. In general the domains of validity of Antoine's laws used in this work include the range of temperatures relevant for Titan's lakes (\textit{i.e.} $90\pm 3$ K). For instance, in the cases of CH$_{4}$ and C$_{2}$H$_{6}$, the lower boundaries are 90.99 K and 91.33 K respectively, implying moderate extrapolations for temperatures slightly below $\sim90$ K.

On the other hand, evaluating the accuracy of Antoine's equations brought by the NIST database is not straightforward. To do so, we have first considered the case of N$_{2}$ for which NIST maintainers derived an Antoine's equation from \cite{edejer_thodos_1967}. These authors published a Frost-Kalkwarf equation based on 180 experimental vapor pressure measurements derived from 13 references. The equation obtained by \cite{edejer_thodos_1967} reproduces the experimental measurements with a deviation ranging between 0.13\% and 2.04\%. Comparing the vapor pressure computed with the NIST Antoine's equation with the one given by \cite{edejer_thodos_1967}, we found differences reaching $\sim 10$\% for the lowest temperatures (\textit{i.e} around 67 K) and $\sim 1$\% for temperatures close to $90$ K. Following a similar approach, we compared pressure computed with the NIST Antoine's equation
and original data from \cite{carruth_kobayashi_1973}. We also made a comparison between our own fit and pressure data given by \cite{handbook74th} for carbon monoxide. For the relevant temperature domain, the deviation
for C$_{2}$H$_{6}$ remains between 0.1\% and 1\%, while the vapor pressure of CO reaches a difference of about 9\%. Consequently
we have fixed the maximum errors on vapor pressures, for all relevant species (\textit{i.e.} N$_{2}$, CH$_{4}$, Ar, CO and C$_{2}$H$_{6}$) to $\pm 10$\% relative to previously used values (see C09). This range should bracket all the vapor pressures expected for each compound. This approach allows us to explore a wide range of possibilities, including combinations which do not correspond to {\bf } physical reality. In this sense, results corresponding to extreme deviations should be regarded as unlikely cases.

Table~\ref{tableMC} gathers our results which are quantified by $\Delta_{P_{\rm vap}}=(X_{\rm max}-X_{\rm min})/\overline{X}$ and the relative standard deviation $\sigma^{*}$ (both expressed in percentage). \Dp measures the total spread (over 10,000 computations of chemical compositions) of mole fraction values for a given species, including the results of the most unlikely combinations of synthetic errors. The relative standard deviation $\sigma^{*}(i) = \sqrt{\overline{X^{2}_{i}}-\overline{X}^{2}_{i}}/\overline{X}_{i}$ (the upper bar denotes the average value over 10,000 computations) shows how much variation or ``dispersion'' there is from the ``average'' $X_{i}$. Table~\ref{tableMC} shows that $\sigma^{*}(i)$'s differ strongly from \Dp's. This feature corresponds to the signature of extremely narrow distributions of values around the average ones and can be explained by the non-linearity of the equations of our model. Indeed, synthetic errors are chosen with {\bf a} uniform distribution but the resulting distribution of mole fractions is heterogeneous. The shapes of $X_{i}$'s distributions are shown in subsection~\ref{allcombi} in which errors for all thermodynamic inputs are taken into consideration.

Errors on vapor pressure mainly affect mole fractions of species for which Equation~\ref{equa1} is written. This behavior is not surprising as Equation~\ref{equa1} contains explicitly the vapor pressure.
The entire set of equations being coupled, even a variation of one vapor pressure affects the {\bf} mole fractions of all the other species determined with our model. Note that the case of ethane is particular because its atmospheric mole fraction $Y_{\rm C_{2}H_{6}}$ is an unknown of our mathematical problem, while atmospheric abundances of N$_{2}$, CH$_{4}$, Ar and CO are fixed by the observations. Compared to other compounds belonging to precipitated species  HCN shows a relatively high $\Delta_{P_{\rm vap}}$ ($\sigma^{*}$). In this case, this is also due to the use of an equation (Equation~\ref{equa2}) where vapor pressures play a role \textit{via} $\Gamma_{i}$. Average values $\overline{X}_{i}$ differ slightly from previous results (C09).
%

\subsection{Influence of molar volume uncertainties}

  Molar volumes have been estimated \textit{via} {\bf } Rackett's method \citep[see][]{poling_2007}. Table 4-11 and pages 4.36--4.37 of \cite{poling_2007} present comparisons between measured molar volumes and estimated ones for some organic compounds at the boiling temperature. 
These comparisons show that maximum deviations typically reach the levels of a few percent. We then adopted a maximum ``error'' of $\pm 10$\% for Monte-Carlo simulations only applied to molar volumes. These simulations allow a sensitivity comparison with those performed for vapor pressure. Resulting \Dv and \Sv are displayed in Table~\ref{tableMC}.

Similarly to simulations related to vapor pressure, large differences between \Dv and \Sv indicate a very narrow spread of mole fractions in lakes. As would be expected, species for which thermodynamic Equations~\ref{equa1}
are explicitly used show the highest deviations.
As also shown by Table~\ref{tableMC}, a general trend is that mole fractions appear to be much more sensitive to molar volume than to vapor pressures.

\subsection{Influence of enthalpies of vaporization uncertainties}

The NIST and the literature provide numerous interpolation formul\ae{ } for enthalpies of vaporization. 
Getting reliable estimates of their actual accuracy is not easy because the domains of validity of these formul\ae{ } do not often include the ground temperature of Titan.
For instance for C$_{2}$H$_{6}$, C$_{3}$H$_{8}$ and C$_{4}$H$_{8}$ extrapolations over about 100 K are required. However we have performed error estimates only for methane, ethane and argon.
In the cases of methane and ethane, we compared the enthalpies of vaporization given by the NIST database (originally published by \cite{majer_svoboda_85}) to those computed with the equations provided by \cite{somayajulu_1988}. For methane, in the temperature range of interest for the surface on Titan, we obtained differences lower than 1\%. In the case of ethane, these differences are much more important and lie between 26\% and 30\%. Considering the work of \cite{tegeler_etal_1999}, we estimate an internal uncertainty of about 1\% for the enthalpy of vaporization of argon. 

 Again, in order to be consistent with others Monte-Carlo simulations, we fixed the maximum ``error'' on the enthalpy of vaporization of each species to $\pm 10$\%, a value which is well within the range of uncertainties found from comparisons. The results are displayed in Table~\ref{tableMC} and show that the induced uncertainties on mole fractions are similar or higher than those obtained for the molar volumes.

\subsection{Influence of $l_{ij}$ uncertainties}

The interaction parameters $l_{ij}$'s represent the interaction between molecules of different species and are essentially determined empirically.
These parameters are fixed to zero in the case of interactions between the same molecules ($\forall i$ $l_{ii} = 0$). 
  In principle these $l_{ij}$ depend on temperature, however for typical
nonpolar mixtures over a modest range of temperature, that dependence is usually small \citep[see][]{poling_2007}.
  As stated by DUB89, they are unknown in many situations. In this work, as 13 species are taken into account in the liquid phase, we need to know 156 parameters ($13 \times 13 -13 = 156$; the 13 $l_{ii}$ being set to zero) and testing the possible influence of each of them in our system does not really make sense. Given the fact that the values of $l_{ij}$ range between $\sim$ 0.02 (DUB89) and 0.09 \citep{poling_2007}, we have performed Monte-Carlo simulations with these parameters set randomly between $0$ and $0.10$ (except $l_{ii} =0$). The sensitivity of mole fractions to $l_{ij}$'s is presented in Table~\ref{tableMC} and appears lower than in the cases of molar volume and enthalpy of vaporization.

\subsection{\label{allcombi}Combination of all thermodynamic uncertainties}

\begin{figure}[h]
\centering
\rotatebox{0}{\resizebox{15.6 cm}{22.0759663865546 cm}{\includegraphics{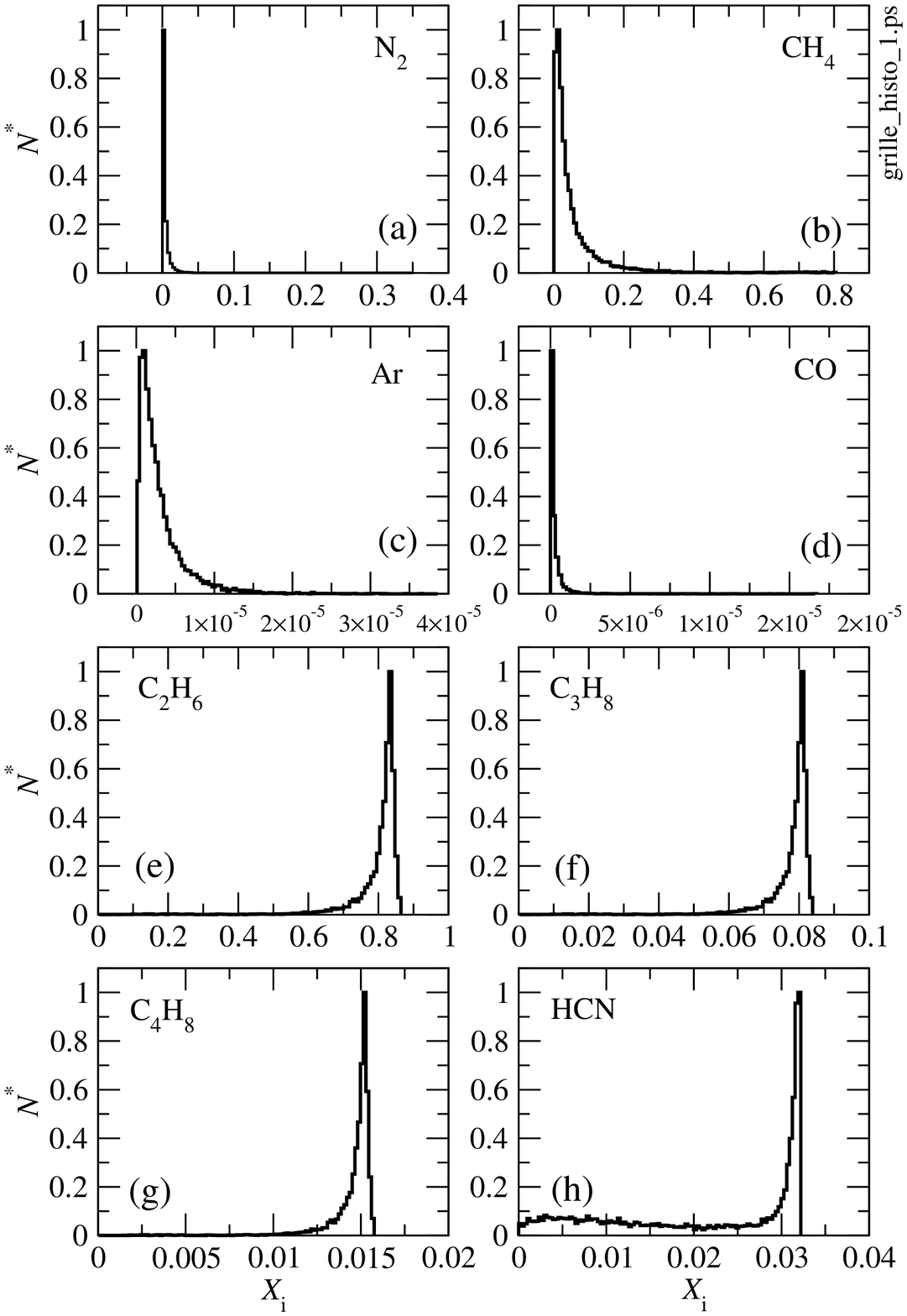}}}
\caption[]{\label{histo1}Histogram of mole fractions of Titan's hydrocarbon lakes. $N^{*}$ is the normalized number of mole fractions owning the value $X_{i}$ computed within a given range of uncertainties for compound $i$ (indicated in the top left of each panel). The 10,000 computations of chemical compositions have been performed at $T= 90$ K, assuming a maximum error of $\pm 10$\% for all $P_{vap}$, $V_{m}$, $\Delta H_{m}$, and assuming a 0--0.1 range of values for the $l_{ij}$'s.}
\end{figure}

\begin{figure}[h]
\centering
\rotatebox{0}{\resizebox{15.6 cm}{18.116974789916 cm}{\includegraphics{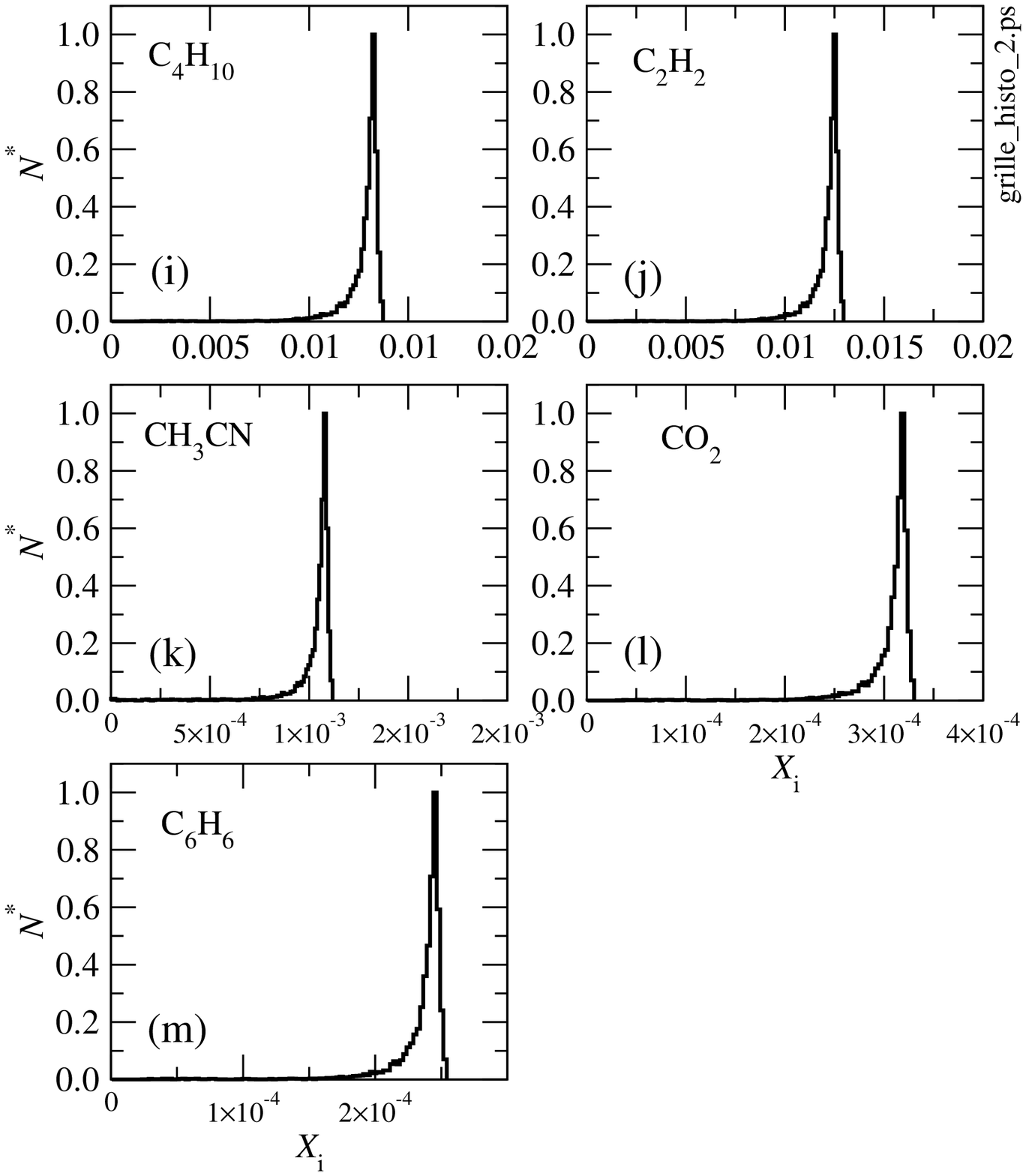}}}
\caption[]{\label{histo2}Same as in Figure~\ref{histo1}.}
\end{figure}

Here we have combined all sources of uncertainties (\textit{i.e.} errors on $P_{vap}$, $V_{m}$, $\Delta H_{vap}$ and
$l_{ij}$) and the results are represented by \DA{ } and \SA{ } in Table~\ref{tableMC}. The combination of uncertainties on thermodynamic data can induce mole fraction fluctuations up to a factor of $\sim 100$ for~\DA~and slightly lower than $\sim 4$ if we consider the relative standard deviation \SA. The distribution of mole fractions $X_{i}$ is represented in the form of histograms in Figures~\ref{histo1} and \ref{histo2}.

For each species, the range $X_{i,min}$ -- $X_{i, max}$ has been divided into 100 intervals. For the $k$-th interval, the number $N_{k}$ of mole fractions owning a given value has been normalized via $N_{k}^{*}= N_{k}/N_{\rm peak}$, where $N_{\rm peak}$ corresponds to the largest $N_{k}$. We stress that $N_{\rm peak}$ has a specific value for each compound. Distributions are very narrow and clearly asymmetric. The smallest abundances are limited by $X=0$. HCN is a particular case for which the highest abundances are limited by the saturation. Curiously, the HCN distribution presents a ``residual tail'' located at lake mole fractions between 0 and $\sim 0.032$.

If we consider {\bf} case 1 of DUB89 (\textit{i.e.} $T= 92.5$ K, $Y_{\rm Ar}=0$ and $Y_{\rm CH_{4}}=0.0155$), our results bracket the
abundances found by these authors. For instance we find $7.7 \times 10^{-6} \le X_{\rm N_{2}} \le 0.039$ 
while DUB89 got $X_{\rm N_{2}} = 0.018$. The case of methane is similar since we find $1.8 \times 10^{-4} \le X_{\rm CH_{4}} \le 0.083$ 
whereas DUB89 inferred $X_{\rm CH_{4}} = 7.3\%$. This illustrates the fact that differences between C09 and DUB89 are consistent with uncertainties caused by poorly known thermodynamic data.

\begin{figure}[h]
\centering
\rotatebox{-90}{\resizebox{12 cm}{16.981512605042 cm}{\includegraphics{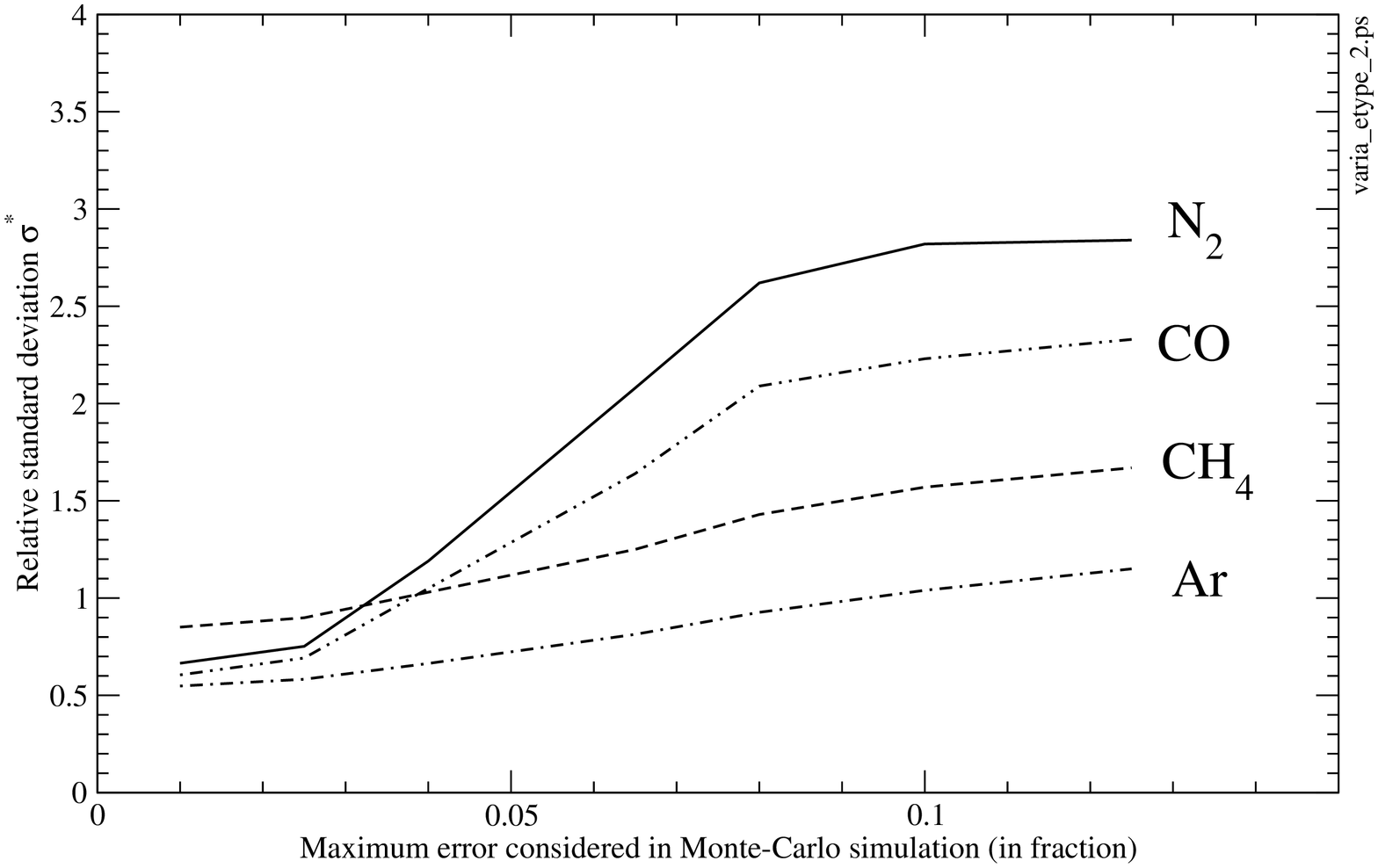}}}
\caption[]{\label{variaET}Relative standard deviation $\sigma^{*}_{i}$ corresponding to the mole fractions of N$_{2}$, CH$_{4}$, Ar and CO. $\sigma^{*}_{i}$ are plotted as functions of the maximum errors obtained for $P_{vap,i}$, $V_{m,i}$, $\delta H_{vap,i}$ in Monte-Carlo simulations. $l_{ij}$'s have been set to 0.1 in all these computations.}
\end{figure}

One could argue that the choice of a maximum deviation of $\pm 10$\% is arbitrary, even if that level of uncertainty has been
discussed in previous subsections. Figure~\ref{variaET} shows the sensitivity of $\sigma^{*}$ to the adopted maximum error for N$_{2}$, CH$_{4}$, Ar and CO. As expected, the standard deviation increases with the value of the maximum error but this behavior appears to be non-linear. During a Monte-Carlo simulation, some combinations of errors yield to a non-convergence of the model, these occurrences corresponding more likely to unphysical situations and/or to an initial input (in practise the nominal solution for $T= 90$ K published in C09) in the Newton-Raphson algorithm which is too far from the solution of the system of the equations. In addition, a maximum error of $\pm 10\%$ in the thermodynamic data is probably an overestimated value even if we do not really know how these data depart from the ``real'' ones. However, we consider that the most important point here was to address the sensitivity of the model to the different sources of uncertainties.

\section{\label{precip}Influence of geographic variations of precipitation}

  On Earth, precipitation is almost entirely water, with the rate dependent on location and time. In the case of Titan, the situation is more complex because the slow sedimentation of stratospheric aerosols to the surface is key to filling the lakes with the dominant photochemical byproducts of methane as well as less abundant species.\\
   Moreover, the sedimentation rates are also a function of location and time.
In this section we restrict our study to the geographic dependence of lakes composition and more precisely to a possible South/North
asymmetry as the distribution of lakes seems to be itself asymmetric \citep[see][]{aharonson_etal_2009}.\\

 As mentioned in Sect.~\ref{model}, the calculations are based on precipitation rates derived from a slightly improved version of \cite{lavvas_etal_2008a,lavvas_etal_2008b} (hereafter LAV08) models that included the atmospheric profile of Titan's atmosphere measured with the Huygens Atmosphere Structure Instrument \citep{fulchignoni_etal_2005} and some updated reaction rates. 
In the following we refer to this set of precipitation rates as LAV09.
    These models are clearly inadequate for a geographical study, this is why we used a 2D models originally developed by 
\cite{lebonnois_etal_2001}.
   These authors used an analytic description of the meridional circulation of Titan's atmosphere to take advection into account in a two-dimensional photochemical model. This coupling between dynamics and photochemistry was subsequently improved by the implementation of the same photochemical model in the IPSL two-dimensional Climate Model \citep[][hereafter LEB08]{crespin_etal_2008}.\\

\begin{figure}[h]
\centering
\rotatebox{-90}{\resizebox{12 cm}{16.981512605042 cm}{\includegraphics{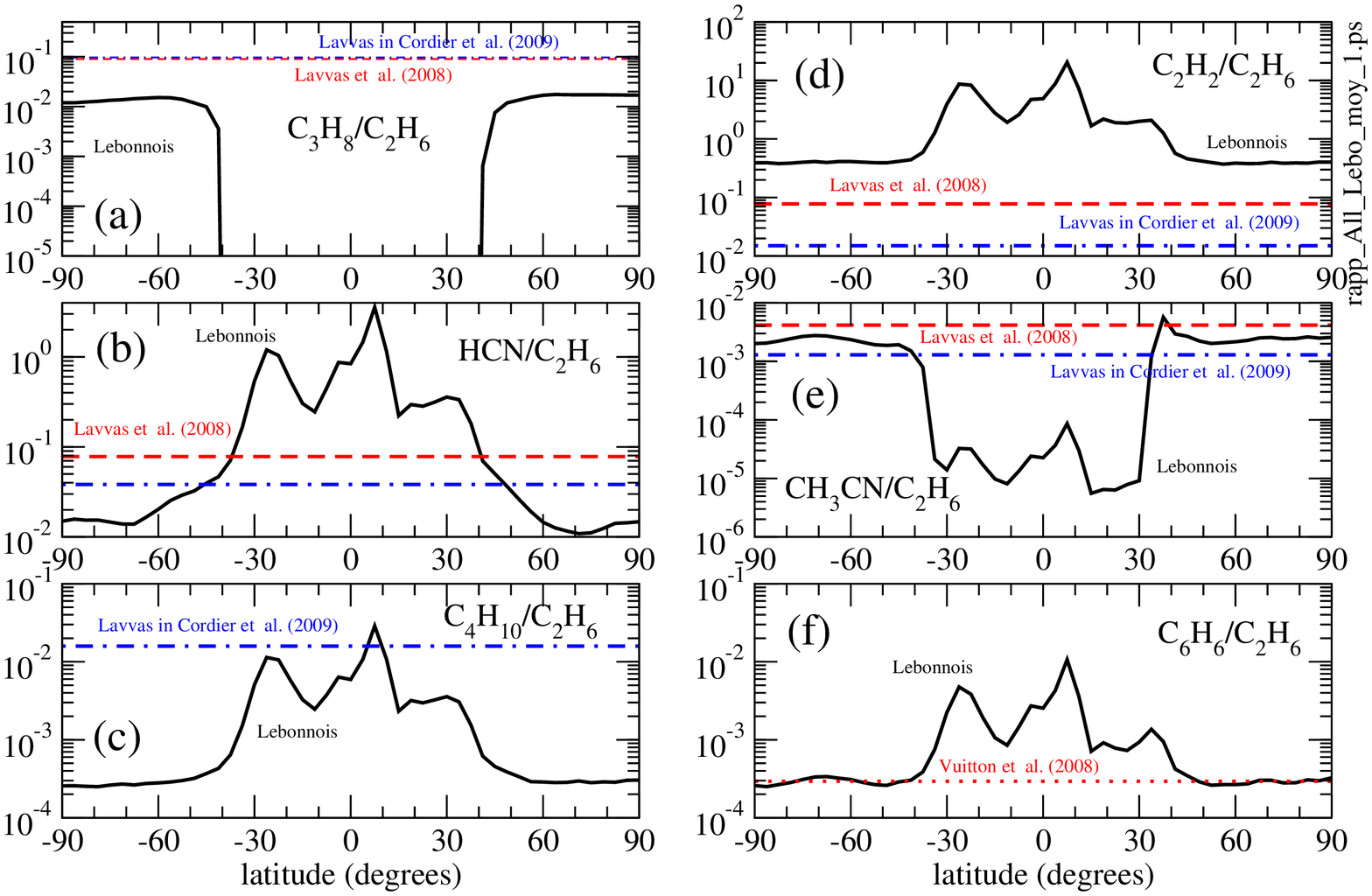}}}
\caption[]{\label{tauxmoyensLebo}Solid lines: time averaged ratios $\tau_{i}/\tau_{\rm C_{2}H_{6}}$ from LEB08 models represented
as a function of Titan's latitude. Dashed lines: the same ratios computed with the LAV09 model and taken from \cite{vuitton_etal_2008} in the case of C$_{6}$H$_{6}$.}
\end{figure}

  Precipitates are included in our thermochemical model via the relation $X_{i} = \frac{\tau_i}{\tau_{C_2H_6}} \times X_{C_2H_6}$. At the first sight, one could believe that multiplying a given $\tau_{i}$ by an arbitrary factor $\alpha_{i}$ yields a multiplication by $\alpha_{i}$ of the resulting mole fraction $X_{i}$. This is not the case because the mole fraction of ethane $X_{\rm C_{2}H_{6}}$ depends on all the other mole fractions. Indeed Eq~\ref{equa1} depends on Eq~\ref{equaphii}, and both equations are solved simultaneously. Hence, the influence of precipitation rates $\tau_{i}$ on mole fractions can only be estimated with a complete calculation. As these rates play a role in our set of equations via the ratios $\tau_{i}/\tau_{\rm C_{2}H_{6}}$, we then consider these ratios instead of absolute $\tau_{i}$ values. For species available in LEB08's models, we computed the time averaged ratios $\tau_{i}/\tau_{\rm C_{2}H_{6}}$, results are displayed in Fig.~\ref{tauxmoyensLebo}, ratios deduced
from LAV08 and LAV09 have been displayed for comparison.

 While we have noticed in LEB08 data absolute precipitation rates have huge latitudinal variations, the ratios $\tau_{i}/\tau_{\rm C_{2}H_{6}}$ do not exhibit such  steep
dependence in regions located poleward of latitudes around $\pm 60^{\rm o}$. In equatorial regions, the precipitation rates computed by LEB08 can be very small, implying that the ratio $\tau_{i}/\tau_{\rm C_{2}H_{6}}$ probably has no great physical meaning in these regions. Fortunately, as hydrocarbon lakes have presumably been detected in polar regions, this observational evidence allows us to identify the questionable values.

   With LEB08 polar ratios $\tau_{i}/\tau_{\rm C_{2}H_{6}}$ we compute two sets of mole fractions: one for the south pole, another for the north
pole. In both cases, the temperature and the pressure have been respectively fixed to 90 K and 1.467 bar. We did not find any
significant differences between the composition of south and north lakes (i.e. differences of the order of 1\%) except for C$_{3}$H$_{8}$ ($\sim 30\%$),
C$_{4}$H$_{10}$, CH$_{3}$CN and C$_{6}$H$_{6}$ (both around $\sim 20\%$). The north pole mole fractions being systematically larger than those computed for
the south pole, this behavior corresponds to LEB08 large ratios at north pole (see Fig.~\ref{tauxmoyensLebo}).
  These results have to be considered carefully because even up-to-data 2D Titan's atmosphere models have to be improved.
For instance, LEB08 precipitation rates are in fact condensation rates. In this approach, when a given species is in an atmospheric layer where the local temperature corresponds to the saturation temperature of this species, then all the molecules in this layer are supposed to precipitate on Titan's surface. In this picture, the microphysics of clouds is not taken into account. Models including this microphysics for a lot
of species have to be developed.

\section{\label{discuss}Discussion and conclusion}
%
%

%
%
  Beside vapor pressures, molar volumes and other parameters already studied in previous sections,
total pressure $P$ and temperature $T$ at ground level could have also an influence on lakes
composition.\\
  The influence of temperature has been already discussed in C09. The variations of ground pressure,
as $P$ appears in Eq.~\ref{equa1}, could change the thermodynamic equilibrium.
Using the Cassini Synthetic Aperture Radar (SAR) 
\cite{stiles_etal_2009} have developed models of the topography of limited portions of the surface, finding surface heights typically
in the range -1500 m to +1000 m, yielding a maximum altitude difference of about
2500 m. More recently, radar altimetry analyzed by \cite{wall_etal_2010} across Ontario Lacus and
its surroundings shows a maximum amplitude of the altimetry echo center of mass of about
$\sim 500$ m. This local determination is compatible with \cite{stiles_etal_2009} work.
HASI data \citep{fulchignoni_etal_2005} contain pressure records, measured on January 14th, 2005
during the Huygens probe descent. Between the Huygens landing site and an altitude of 2500 m,
the pressure ranges between 1467 hPa and 1296 hPa, \textit{i.e.} a relative variation of 10\%.
By means of their 3D general circulation model, \cite{tokano_neubauer_2002} have investigated the
influence of Saturn's gravitational tide on the atmosphere of Titan. They found that induced 
surface pressure variations remains lower than 1.5 hPa, \textit{i.e.} 0.1\%, which is negligible in
our context. Table~\ref{influP} summarizes our results when we change the pressure
value by $\pm 10$\% from the \textit{Huygens} Huygens-derived values $P_{\rm Huy}$.\\

%
   It is a valid question as to whether thermodynamic equilibrium is a reasonable assumption or not. For that
purpose one can calculate the thermal relaxation time of a lake
 with a depth of $H$ given by
\begin{equation}
  \tau \sim H^{2}/\chi
\end{equation}
\citep[see the classical textbook][]{landau_lifshitz_1987}, with $\chi$ the thermal diffusivity which is
given by $\chi = \kappa/\rho C_{\rm P}$ where $\kappa$ (W.m$^{-1}$.K$^{-1}$) is the thermal conductivity,
$\rho$ (kg.m$^{-3}$) the density and $C_{\rm P}$ (J.kg$^{-1}$.K$^{-1}$) the mass,
 specific heat capicity at
constant pressure. We estimated $\kappa$ thanks to the Enskog's theory \citep[see][]{dymond_1985}, we found
$\kappa \sim 0.59$ W.m$^{-1}$.K$^{-1}$ for pure liquid ethane at $T = 90$ K. This value is of the same order of
magnitude as that found by \cite{lorenz_etal_2010} (see their table 1), who got $\kappa \sim 0.25$ W.m$^{-1}$.K$^{-1}$.
Finally we obtained $\tau \sim 2$ Titan's days for $H = 1$ m. As it can be seen in \cite{tokano_2005} 
Titan's surface temperature variations in the polar regions are of the order of 1-2 K over a Titan's year
(about 673 Titan's days\footnote{One Titan's day corresponds approximately to 16 terrestrial days.}), 
that means that --at least the first meter of the lakes-- have enough time to be
in thermal equilibrium with the atmosphere. Following the models of \cite{tokano_2005}, during spring and summer
the lakes are thermally stratified (which may imply a stratified chemical composition), but by the autumnal equinox 
convection renders the upper layer of the lakes isothermal.\\

%
       In our work we use the semi-empirical regular solution theory already employed in previous works
\citep{dubouloz_etal_1989,cordier_etal_2009,cordier_etal_2010} but this also has its limits of validity. 
Particularly, it has been introduced for binary mixtures of nonpolar molecules \citep{hildebrand_scott_1962}
and generalized to multicomponent mixtures \citep[see][chaper 8]{poling_2007}. Even though this generalization
appears to be reasonable it has never been properly validated in a context relevant for Titan's hydrocarbon
lakes. Moreover the interaction parameters (the $l_{ij}$'s) are not well known in the case of nonpolar
molecules and are probably an inadequate formalism when polar components are in the solution. We recall that the mixture
considered in this work includes two polar molecules: HCN (with a dipole moment of 2.98 D, which could be compared to the water
dipole moment of 1.85 D) and acetonitrile CH$_{3}$CN (with a dipole moment of 3.84 D).\\

%
   Our work stresses the great impact on predicted composition of uncertainties in the thermodynamic inputs. In the framework of our model,
this influence appears to be more important than abundance differences between north and south pole lakes, assuming
a maximum ``error'' level of $\pm 10$\% considered in our Monte-Carlo simulations.
Our computations show also that the influence of pressure variations is purely negligible.\\
  It is important to note that, in this work, we did not consider the temporal variation of lakes chemical composition. Indeed, many phenomena
could contribute to these variations, among which seasonal variations for short timescales and the Milankovitch cycle
for longer timescales \citep[see][]{aharonson_etal_2009}. We stress that all these phenomena involve processes of evaporation/condensation of various species (in particular CH$_4$), which clearly represent non-equilibrium situations. A more realistic model will have to take into account
energy and mass fluxes between lakes and the atmosphere \citep[see][]{tokano_2005} and incorporate a chemical model similar to the one used
in the present work. Future works will have to integrate these two aspects of modelling to provide a more accurate description of
lakes evolution.

    We also underline the need for more realistic photochemistry models as already stated by \citep{hebrard_etal_2007} among others.
   If we concentrate on the properties of the liquid of the lakes themselves, two kinds of approaches can be considered to make progress beyond what has been done here: 
(1) the development of more accurate thermodynamic data (measured in dedicated
experiments and/or determined by \textit{ab initio} computations); (2) Titan's lakes \textit{in vitro} simulations, in which one explicitly attempts to simulate Titan's lakes through a liquid hydrocarbon mixture in contact with an atmosphere in a laboratory chamber. The second approach is surely more relevant because a model is not required to apply pure thermodynamic data, but the first approach may be more practical in the absence of a major experimental effort tied to proposed future missions to Titan like \textit{Titan Saturn System Mission} \citep[TSSM, see][]{matson_etal_2009} or
\textit{Titan Mare Explorer} \citep[TiME, see][]{stofan_etal_2010}, this latter being dedicated to lakes study and analysis.


\bibliographystyle{elsarticle-harv}

\begin{thebibliography}{41}
\expandafter\ifx\csname natexlab\endcsname\relax\def\natexlab#1{#1}\fi
\expandafter\ifx\csname url\endcsname\relax
  \def\url#1{\texttt{#1}}\fi
\expandafter\ifx\csname urlprefix\endcsname\relax\def\urlprefix{URL }\fi

\bibitem[{{Aharonson} et~al.(2009){Aharonson}, {Hayes}, {Lunine}, {Lorenz},
  {Allison}, and {Elachi}}]{aharonson_etal_2009}
{Aharonson}, O., {Hayes}, A.~G., {Lunine}, J.~I., {Lorenz}, R.~D., {Allison},
  M.~D., {Elachi}, C., Dec. 2009. {An asymmetric distribution of lakes on Titan
  as a possible consequence of orbital forcing}. Nature Geoscience 2, 851--854.

\bibitem[{{Brown} et~al.(2008){Brown}, {Soderblom}, {Soderblom}, {Clark},
  {Jaumann}, {Barnes}, {Sotin}, {Buratti}, {Baines}, and
  {Nicholson}}]{brown_etal_2008}
{Brown}, R.~H., {Soderblom}, L.~A., {Soderblom}, J.~M., {Clark}, R.~N.,
  {Jaumann}, R., {Barnes}, J.~W., {Sotin}, C., {Buratti}, B., {Baines}, K.~H.,
  {Nicholson}, P.~D., Jul. 2008. {The identification of liquid ethane in
  Titan's Ontario Lacus}. Nature 454, 607--610.

\bibitem[{{Carruth} and {Kobayashi}(1973)}]{carruth_kobayashi_1973}
{Carruth}, G.~F., {Kobayashi}, R., 1973. {Vapor Pressure of Normal Paraffins
  Ethane Through n-Decane from Their Triple Points to About 10 Mm Hg}. J. Chem.
  Eng. Data 18, 115--126.

\bibitem[{{Cordier} et~al.(2009){Cordier}, {Mousis}, {Lunine}, {Lavvas}, and
  {Vuitton}}]{cordier_etal_2009}
{Cordier}, D., {Mousis}, O., {Lunine}, J.~I., {Lavvas}, P., {Vuitton}, V., Dec.
  2009. {An Estimate of the Chemical Composition of Titan's Lakes}. ApJL 707,
  L128--L131.

\bibitem[{{Cordier} et~al.(2010){Cordier}, {Mousis}, {Lunine}, {Lebonnois},
  {Lavvas}, {Lobo}, and {Ferreira}}]{cordier_etal_2010}
{Cordier}, D., {Mousis}, O., {Lunine}, J.~I., {Lebonnois}, S., {Lavvas}, P.,
  {Lobo}, L.~Q., {Ferreira}, A.~G.~M., Aug. 2010. {About the possible role of
  hydrocarbon lakes in the origin of Titan's noble gas atmospheric depletion}.
  ArXiv e-prints.

\bibitem[{{Crespin} et~al.(2008){Crespin}, {Lebonnois}, {Vinatier},
  {B{\'e}zard}, {Coustenis}, {Teanby}, {Achterberg}, {Rannou}, and
  {Hourdin}}]{crespin_etal_2008}
{Crespin}, A., {Lebonnois}, S., {Vinatier}, S., {B{\'e}zard}, B., {Coustenis},
  A., {Teanby}, N.~A., {Achterberg}, R.~K., {Rannou}, P., {Hourdin}, F., Oct.
  2008. {Diagnostics of Titan's stratospheric dynamics using Cassini/CIRS data
  and the 2-dimensional IPSL circulation model}. Icarus 197, 556--571.

\bibitem[{{Dubouloz} et~al.(1989){Dubouloz}, {Raulin}, {Lellouch}, and
  {Gautier}}]{dubouloz_etal_1989}
{Dubouloz}, N., {Raulin}, F., {Lellouch}, E., {Gautier}, D., Nov. 1989.
  {Titan's hypothesized ocean properties - The influence of surface temperature
  and atmospheric composition uncertainties}. Icarus 82, 81--96.

\bibitem[{{Dymond}(1985)}]{dymond_1985}
{Dymond}, J.~H., 1985. {Hard-sphere Theories of Transport Properties}. Chem.
  Soc. Rev. 14, 317--356.

\bibitem[{{Edejer} and {Thodos}(1967)}]{edejer_thodos_1967}
{Edejer}, M.~R., {Thodos}, G., 1967. J. Chem. Eng. Data 12, 206--209.

\bibitem[{{Flasar}(1983)}]{flasar_1983}
{Flasar}, F.~M., Jul. 1983. {Oceans on Titan?} Science 221, 55--57.

\bibitem[{{Fulchignoni} et~al.(2005){Fulchignoni}, {Ferri}, {Angrilli}, {Ball},
  {Bar-Nun}, {Barucci}, {Bettanini}, {Bianchini}, {Borucki}, {Colombatti},
  {Coradini}, {Coustenis}, {Debei}, {Falkner}, {Fanti}, {Flamini}, {Gaborit},
  {Grard}, {Hamelin}, {Harri}, {Hathi}, {Jernej}, {Leese}, {Lehto}, {Lion
  Stoppato}, {L{\'o}pez-Moreno}, {M{\"a}kinen}, {McDonnell}, {McKay},
  {Molina-Cuberos}, {Neubauer}, {Pirronello}, {Rodrigo}, {Saggin},
  {Schwingenschuh}, {Seiff}, {Sim{\~o}es}, {Svedhem}, {Tokano}, {Towner},
  {Trautner}, {Withers}, and {Zarnecki}}]{fulchignoni_etal_2005}
{Fulchignoni}, M., {Ferri}, F., {Angrilli}, F., {Ball}, A.~J., {Bar-Nun}, A.,
  {Barucci}, M.~A., {Bettanini}, C., {Bianchini}, G., {Borucki}, W.,
  {Colombatti}, G., {Coradini}, M., {Coustenis}, A., {Debei}, S., {Falkner},
  P., {Fanti}, G., {Flamini}, E., {Gaborit}, V., {Grard}, R., {Hamelin}, M.,
  {Harri}, A.~M., {Hathi}, B., {Jernej}, I., {Leese}, M.~R., {Lehto}, A., {Lion
  Stoppato}, P.~F., {L{\'o}pez-Moreno}, J.~J., {M{\"a}kinen}, T., {McDonnell},
  J.~A.~M., {McKay}, C.~P., {Molina-Cuberos}, G., {Neubauer}, F.~M.,
  {Pirronello}, V., {Rodrigo}, R., {Saggin}, B., {Schwingenschuh}, K., {Seiff},
  A., {Sim{\~o}es}, F., {Svedhem}, H., {Tokano}, T., {Towner}, M.~C.,
  {Trautner}, R., {Withers}, P., {Zarnecki}, J.~C., Dec. 2005. {In situ
  measurements of the physical characteristics of Titan's environment}. Nature
  438, 785--791.

\bibitem[{{Giauque} and {Ruehrwein}(1939)}]{giauque_ruehrwein_1939}
{Giauque}, W.~F., {Ruehrwein}, R.~A., 1939. {The Entropy of Hydrogen Cyanide.
  Heat Capacity, Heat of Vaporization and Vapor Pressure. Hydrogen Bond
  Polymerization of the Gas in Chains of Indefinite Length}. J. Am. Chem. Soc.
  61, 2626.

\bibitem[{{Hayes} et~al.(2010){Hayes}, {Wolf}, {Aharonson}, {Zebker}, {Lorenz},
  {Kirk}, {Paillou}, {Lunine}, {Wye}, {Callahan}, {Wall}, and
  {Elachi}}]{hayes_etal_2010}
{Hayes}, A.~G., {Wolf}, A.~S., {Aharonson}, O., {Zebker}, H., {Lorenz}, R.,
  {Kirk}, R.~L., {Paillou}, P., {Lunine}, J., {Wye}, L., {Callahan}, P.,
  {Wall}, S., {Elachi}, C., Sep. 2010. {Bathymetry and absorptivity of Titan's
  Ontario Lacus}. Journal of Geophysical Research (Planets) 115, 9009--+.

\bibitem[{{H{\'e}brard} et~al.(2007){H{\'e}brard}, {Dobrijevic}, {B{\'e}nilan},
  and {Raulin}}]{hebrard_etal_2007}
{H{\'e}brard}, E., {Dobrijevic}, M., {B{\'e}nilan}, Y., {Raulin}, F., Jul.
  2007. {Photochemical kinetics uncertainties in modeling Titan's atmosphere:
  First consequences}. PSS 55, 1470--1489.

\bibitem[{{Hildebrand} and {Scott}(1962)}]{hildebrand_scott_1962}
{Hildebrand}, J.~H., {Scott}, R.~L., 1962. {Regular Solutions}. Prentice-Hall,
  Englewood Cliffs, N.J.

\bibitem[{{Jennings} et~al.(2009){Jennings}, {Flasar}, {Kunde}, {Samuelson},
  {Pearl}, {Nixon}, {Carlson}, {Mamoutkine}, {Brasunas}, {Guandique},
  {Achterberg}, {Bjoraker}, {Romani}, {Segura}, {Albright}, {Elliott},
  {Tingley}, {Calcutt}, {Coustenis}, and {Courtin}}]{jennings_etal_2009}
{Jennings}, D.~E., {Flasar}, F.~M., {Kunde}, V.~G., {Samuelson}, R.~E.,
  {Pearl}, J.~C., {Nixon}, C.~A., {Carlson}, R.~C., {Mamoutkine}, A.~A.,
  {Brasunas}, J.~C., {Guandique}, E., {Achterberg}, R.~K., {Bjoraker}, G.~L.,
  {Romani}, P.~N., {Segura}, M.~E., {Albright}, S.~A., {Elliott}, M.~H.,
  {Tingley}, J.~S., {Calcutt}, S., {Coustenis}, A., {Courtin}, R., Feb. 2009.
  {Titan's Surface Brightness Temperatures}. ApJL 691, L103--L105.

\bibitem[{{Landau} and {Lifshitz}(1987)}]{landau_lifshitz_1987}
{Landau}, L.~D., {Lifshitz}, E.~M., 1987. {Fluid Mechanics (Course of
  Theoretical Physics)}, 2nd Edition. Butterworth-Heinemann.

\bibitem[{{Lavvas} et~al.(2008{\natexlab{a}}){Lavvas}, {Coustenis}, and
  {Vardavas}}]{lavvas_etal_2008a}
{Lavvas}, P.~P., {Coustenis}, A., {Vardavas}, I.~M., Jan. 2008{\natexlab{a}}.
  {Coupling photochemistry with haze formation in Titan's atmosphere, Part I:
  Model description}. Planet. Space Sci. 56, 27--66.

\bibitem[{{Lavvas} et~al.(2008{\natexlab{b}}){Lavvas}, {Coustenis}, and
  {Vardavas}}]{lavvas_etal_2008b}
{Lavvas}, P.~P., {Coustenis}, A., {Vardavas}, I.~M., Jan. 2008{\natexlab{b}}.
  {Coupling photochemistry with haze formation in Titan's atmosphere, Part II:
  Results and validation with Cassini/Huygens data}. Planet. Space Sci. 56,
  67--99.

\bibitem[{{Lebonnois} et~al.(2001){Lebonnois}, {Toublanc}, {Hourdin}, and
  {Rannou}}]{lebonnois_etal_2001}
{Lebonnois}, S., {Toublanc}, D., {Hourdin}, F., {Rannou}, P., Aug. 2001.
  {Seasonal Variations of Titan's Atmospheric Composition}. Icarus 152,
  384--406.

\bibitem[{Lide(1974)}]{handbook74th}
Lide, D.~P. (Ed.), 1974. CRC Handbook of Chemistry and Physics, 74th Edition.
  CRC PRESS.

\bibitem[{{Lorenz} et~al.(2003){Lorenz}, {Biolluz}, {Encrenaz}, {Janssen},
  {West}, and {Muhleman}}]{lorenz_etal_2003}
{Lorenz}, R.~D., {Biolluz}, G., {Encrenaz}, P., {Janssen}, M.~A., {West},
  R.~D., {Muhleman}, D.~O., Apr. 2003. {Cassini RADAR: prospects for Titan
  surface investigations using the microwave radiometer}. PSS 51, 353--364.

\bibitem[{{Lorenz} et~al.(2010){Lorenz}, {Newman}, and
  {Lunine}}]{lorenz_etal_2010}
{Lorenz}, R.~D., {Newman}, C., {Lunine}, J.~I., Jun. 2010. {Threshold of wave
  generation on Titan's lakes and seas: Effect of viscosity and implications
  for Cassini observations}. Icarus 207, 932--937.

\bibitem[{{Lunine} et~al.(1983){Lunine}, {Stevenson}, and
  {Yung}}]{lunine_etal_1983}
{Lunine}, J.~I., {Stevenson}, D.~J., {Yung}, Y.~L., Dec. 1983. {Ethane ocean on
  Titan}. Science 222, 1229--+.

\bibitem[{{Majer} and {Svoboda}(1985)}]{majer_svoboda_85}
{Majer}, V., {Svoboda}, V., 1985. {Enthalpies of vaporization of organic
  compounds: A critical Review and Data Compilation}. Blackwell Scientific
  Publications, Oxford.

\bibitem[{{Matson} et~al.(2009){Matson}, {Coustenis}, {Lunine}, {Lebreton},
  {Reh}, {Beauchamp}, and {Erd}}]{matson_etal_2009}
{Matson}, D., {Coustenis}, A., {Lunine}, J.~I., {Lebreton}, J., {Reh}, K.,
  {Beauchamp}, P., {Erd}, C., Dec. 2009. {Spacecraft Exploration of Titan and
  Enceladus}. AGU Fall Meeting Abstracts, D1474+.

\bibitem[{{McKay} et~al.(1993){McKay}, {Pollack}, {Lunine}, and
  {Courtin}}]{mckay_etal_1993}
{McKay}, C.~P., {Pollack}, J.~B., {Lunine}, J.~I., {Courtin}, R., Mar. 1993.
  {Coupled atmosphere-ocean models of Titan's past}. Icarus 102, 88--98.

\bibitem[{{Metropolis} and {Ulam}(1949)}]{metropolis_ulam_1949}
{Metropolis}, N., {Ulam}, S., 1949. J. Amer. Statistical Assoc. 44, 335.

\bibitem[{{Muhleman} et~al.(1995){Muhleman}, {Grossman}, and
  {Butler}}]{muhleman_etal_1995}
{Muhleman}, D.~O., {Grossman}, A.~W., {Butler}, B.~J., 1995. {Radar
  Investigations of Mars, Mercury, and Titan}. Annual Review of Earth and
  Planetary Sciences 23, 337--374.

\bibitem[{{Niemann} et~al.(2005){Niemann}, {Atreya}, {Bauer}, {Carignan},
  {Demick}, {Frost}, {Gautier}, {Haberman}, {Harpold}, {Hunten}, {Israel},
  {Lunine}, {Kasprzak}, {Owen}, {Paulkovich}, {Raulin}, {Raaen}, and
  {Way}}]{niemann_etal_2005}
{Niemann}, H.~B., {Atreya}, S.~K., {Bauer}, S.~J., {Carignan}, G.~R., {Demick},
  J.~E., {Frost}, R.~L., {Gautier}, D., {Haberman}, J.~A., {Harpold}, D.~N.,
  {Hunten}, D.~M., {Israel}, G., {Lunine}, J.~I., {Kasprzak}, W.~T., {Owen},
  T.~C., {Paulkovich}, M., {Raulin}, F., {Raaen}, E., {Way}, S.~H., Dec. 2005.
  {The abundances of constituents of Titan's atmosphere from the GCMS
  instrument on the Huygens probe}. Nature 438, 779--784.

\bibitem[{{Poling} et~al.(2007){Poling}, {Prausnitz}, and
  {O'Connell}}]{poling_2007}
{Poling}, B.~E., {Prausnitz}, J.~M., {O'Connell}, J., 2007. {The Properties of
  Gases and Liquids}, 5th Edition. McGraw-Hill Professional, Englewood Cliffs.

\bibitem[{{Somayajulu}(1988)}]{somayajulu_1988}
{Somayajulu}, G.~R., 1988. Int. J. Thermophys. 9, 567.

\bibitem[{{Stiles} et~al.(2009){Stiles}, {Hensley}, {Gim}, {Bates}, {Kirk},
  {Hayes}, {Radebaugh}, {Lorenz}, {Mitchell}, {Callahan}, {Zebker}, {Johnson},
  {Wall}, {Lunine}, {Wood}, {Janssen}, {Pelletier}, {West}, {Veeramacheneni},
  and {Cassini RADAR Team}}]{stiles_etal_2009}
{Stiles}, B.~W., {Hensley}, S., {Gim}, Y., {Bates}, D.~M., {Kirk}, R.~L.,
  {Hayes}, A., {Radebaugh}, J., {Lorenz}, R.~D., {Mitchell}, K.~L., {Callahan},
  P.~S., {Zebker}, H., {Johnson}, W.~T.~K., {Wall}, S.~D., {Lunine}, J.~I.,
  {Wood}, C.~A., {Janssen}, M., {Pelletier}, F., {West}, R.~D.,
  {Veeramacheneni}, C., {Cassini RADAR Team}, Aug. 2009. {Determining Titan
  surface topography from Cassini SAR data}. Icarus 202, 584--598.

\bibitem[{{Stofan} et~al.(2007){Stofan}, {Elachi}, {Lunine}, {Lorenz},
  {Stiles}, {Mitchell}, {Ostro}, {Soderblom}, {Wood}, {Zebker}, {Wall},
  {Janssen}, {Kirk}, {Lopes}, {Paganelli}, {Radebaugh}, {Wye}, {Anderson},
  {Allison}, {Boehmer}, {Callahan}, {Encrenaz}, {Flamini}, {Francescetti},
  {Gim}, {Hamilton}, {Hensley}, {Johnson}, {Kelleher}, {Muhleman}, {Paillou},
  {Picardi}, {Posa}, {Roth}, {Seu}, {Shaffer}, {Vetrella}, and
  {West}}]{stofan_etal_2007}
{Stofan}, E.~R., {Elachi}, C., {Lunine}, J.~I., {Lorenz}, R.~D., {Stiles}, B.,
  {Mitchell}, K.~L., {Ostro}, S., {Soderblom}, L., {Wood}, C., {Zebker}, H.,
  {Wall}, S., {Janssen}, M., {Kirk}, R., {Lopes}, R., {Paganelli}, F.,
  {Radebaugh}, J., {Wye}, L., {Anderson}, Y., {Allison}, M., {Boehmer}, R.,
  {Callahan}, P., {Encrenaz}, P., {Flamini}, E., {Francescetti}, G., {Gim}, Y.,
  {Hamilton}, G., {Hensley}, S., {Johnson}, W.~T.~K., {Kelleher}, K.,
  {Muhleman}, D., {Paillou}, P., {Picardi}, G., {Posa}, F., {Roth}, L., {Seu},
  R., {Shaffer}, S., {Vetrella}, S., {West}, R., Jan. 2007. {The lakes of
  Titan}. Nature 445, 61--64.

\bibitem[{{Stofan} et~al.(2010){Stofan}, {Lunine}, and
  {Lorenz}}]{stofan_etal_2010}
{Stofan}, E.~R., {Lunine}, J., {Lorenz}, R., Apr. 2010. {The lakes and seas of
  Titan: outstanding questions and future exploration}. In: {V.~Cottini,
  C.~Nixon, \& R.~Lorenz} (Ed.), Through Time; A Workshop On Titan's Past,
  Present and Future. pp. 48--+.

\bibitem[{{Tegeler} et~al.(1999){Tegeler}, {Span}, and
  {Wagner}}]{tegeler_etal_1999}
{Tegeler}, C., {Span}, S., {Wagner}, W., 1999. {A New Equation of State for
  Argon Covering the Fluid Region for Temperatures From the Melting Line to 700
  K at Pressures up to 1000 MPa}. J. Phys. Chem. Ref. Data 779, 829.

\bibitem[{{Tokano}(2005)}]{tokano_2005}
{Tokano}, T., 2005. {Thermal structure of putative hydrocarbon lakes on Titan}.
  Advances in Space Research 36, 286--294.

\bibitem[{{Tokano} and {Neubauer}(2002)}]{tokano_neubauer_2002}
{Tokano}, T., {Neubauer}, F.~M., Aug. 2002. {Tidal Winds on Titan Caused by
  Saturn}. Icarus 158, 499--515.

\bibitem[{{Turtle} et~al.(2007){Turtle}, {Perry}, {McEwen}, {West}, {Dawson},
  {Porco}, and {Fussner}}]{turtle_etal_2007}
{Turtle}, E.~P., {Perry}, J.~E., {McEwen}, A.~S., {West}, R.~A., {Dawson},
  D.~D., {Porco}, C.~C., {Fussner}, S., Aug. 2007. {Cassini Imaging Science
  Subsystem Observations of Titan's High-Latitude Lakes}. LPI Contributions
  1357, 142--143.

\bibitem[{{Vuitton} et~al.(2008){Vuitton}, {Yelle}, and
  {Cui}}]{vuitton_etal_2008}
{Vuitton}, V., {Yelle}, R.~V., {Cui}, J., May 2008. {Formation and distribution
  of benzene on Titan}. Journal of Geophysical Research (Planets) 113, 5007--+.

\bibitem[{{Wall} et~al.(2010){Wall}, {Hayes}, {Bristow}, {Lorenz}, {Stofan},
  {Le Gall}, {Janssen}, {Lopes}, {Wye}, {Soderblom}, {Paillou}, {Aharonson},
  {Zebker}, {Farr}, {Mitri}, {Kirk}, {Mitchell}, {Notarnicola}, {Casarano}, and
  {Ventura}}]{wall_etal_2010}
{Wall}, S., {Hayes}, A., {Bristow}, C., {Lorenz}, R., {Stofan}, E., L.~J., {Le
  Gall}, A., {Janssen}, M., {Lopes}, R., {Wye}, L., {Soderblom}, L., {Paillou},
  P., {Aharonson}, O., {Zebker}, H., {Farr}, T., {Mitri}, R., {Kirk}, R.,
  {Mitchell}, K., {Notarnicola}, C., {Casarano}, D., {Ventura}, B., 2010.
  {Active shoreline of Ontario Lacus, Titan: A morphological study of the lake
  and its surroundings}. Geophysical Research Letters 37, L05202.

\end{thebibliography}

\newpage
\begin{table}[h]
\centering
\caption{Assumed composition of Titan's atmosphere at the ground level.}
\begin{tabular}{lcc}
\hline
\hline
\noalign{\smallskip}
Atmosphere            & Mole fraction                        & Determination                                 \\
\hline
H$_2$                & $9.8 \times 10^{-4}$            & Huygens GCMS$^{(a)}$ 
\\
CH$_4$                & 0.0492                        & Huygens GCMS$^{(b)}$                    \\
CO                    & $4.70 \times 10^{-5}$            & Cassini CIRS$^{(c)}$                          \\
$^{40}$Ar                & $4.32 \times 10^{-5}$                & Huygens GCMS$^{(b)}$                    \\
N$_2$                & 0.95                                      & C09$^{(d)}$                            \\
C$_2$H$_6$                  & $1.49 \times 10^{-5}$                 & C09$^{(d)}$                            \\

\hline
\end{tabular}
\newline
$^{(a)}$Owen \& Niemann 2009; $^{(b)}$Niemann et al. 2005; $^{(c)}$De Kok et al. 2007; $^{(d)}$N$_2$ and C$_2$H$_6$ abundances have been calculated by C09 and correspond to a ground temperature of 93.65 K. In Monte-Carlo simulations presented here, the mole fractions of N$_2$ and C$_2$H$_6$ are  varying (see text).
\label{atmoscompo}
\end{table}
\newpage
\begin{sidewaystable}
\caption[]{Results of Monte-Carlo simulations showing the space of possible values for thermodynamic inputs explored randomly within a $\pm 10$\% range centered on the nominal values used by C09. $X_{\rm min}$, $\overline{X}$ and $X_{\rm max}$ are recorded for each species and results are presented in the form of $\Delta$'s and relative standard deviations $\sigma^{*}$ (see text). $X_{\rm min}$, $\overline{X}$ and $X_{\rm max}$ are shown only in the cases where the space of vapor pressures is explored. \Dp (\Sp) refers to Monte-Carlo simulations for which only vapor pressures are affected by synthetic errors. \Dv (\Sv), \Dh (\Sh), and \Dl (\Sl) refer to molar volumes, enthalpies of vaporization and interaction parameters $l_{ij}$, respectively. \DA (\SA) are resulting uncertainties when all thermodynamic quantities are considered with errors.}
\begin{center}
{
\begin{tabular}{lcccrr|rrrrrr|rr}
\hline
\hline
               & $X_{\rm min}$     & $\overline{X} $     & $X_{\rm max}$     &  \Dp  & \Sp &  \Dv & \Sv & \Dh  & \Sh &   \Dl & \Sl & \DA  & \SA \\
               & $(P_{\rm vap})$   & $(P_{\rm vap})$  & $(P_{\rm vap})$   & (\%)  & (\%)& (\%) & (\%)& (\%) & (\%)&  (\%) & (\%)& (\%) & (\%)\\ 
\hline
N$_{2}$         &   $4.1\tento{-3}$ &   $4.9\tento{-3}$ &   $5.9\tento{-3}$ &  37   & 8 &   792 & 94 & 2310  & 105 &  445  &  65 & 8540 & 282 \\
CH$_{4}$        &   $8.3\tento{-2}$ &   $9.7\tento{-2}$ &   $1.1\tento{-1}$ &  32   & 9 &   346 & 61 &  570  &  63 &  379  &  84 & 1370 & 157 \\
Ar              &   $4.3\tento{-6}$ &   $4.9\tento{-6}$ &   $5.8\tento{-6}$ &  31   & 7 &   333 & 53 &  557  &  56 &  288  &  54 & 1270 & 104 \\
CO              &   $3.5\tento{-7}$ &   $4.2\tento{-7}$ &   $5.1\tento{-7}$ &  37   & 8 &   734 & 88 &  244  & 102 &  386  &  59 & 5840 & 223 \\
C$_{2}$H$_{6}$  &   $7.5\tento{-1}$ &   $7.6\tento{-1}$ &   $7.7\tento{-1}$ &   3   & 1 &    50 &  8 &   96  &   9 &   17  &   4 &  108 &  10 \\
C$_{3}$H$_{8}$  &   $7.3\tento{-2}$ &   $7.4\tento{-2}$ &   $7.5\tento{-2}$ &   3   & 1 &    50 &  8 &   96  &   9 &   17  &   4 &  108 &  10 \\
C$_{4}$H$_{8}$  &   $1.4\tento{-2}$ &   $1.4\tento{-2}$ &   $1.4\tento{-2}$ &   3   & 1 &    50 &  8 &   96  &   9 &   17  &   4 &  108 &  10 \\
HCN             &   $2.0\tento{-2}$ &   $2.2\tento{-2}$ &   $2.3\tento{-2}$ &  16   & 4 &   156 & 50 &  167  &  56 &   52  &  13 &  150 &  52 \\
C$_{4}$H$_{10}$ &   $1.2\tento{-2}$ &   $1.2\tento{-2}$ &   $1.2\tento{-2}$ &   3   & 1 &    50 &  8 &   96  &   9 &   17  &   4 &  108 &  10 \\
C$_{2}$H$_{2}$  &   $1.1\tento{-2}$ &   $1.1\tento{-2}$ &   $1.2\tento{-2}$ &   3   & 1 &    50 &  8 &   96  &   9 &   17  &   4 &  108 &  10 \\
CH$_{3}$CN      &   $9.7\tento{-4}$ &   $9.9\tento{-4}$ &   $1.0\tento{-3}$ &   3   & 1 &    52 &  8 &  112  &  11 &   17  &   4 &  109 &  14 \\
CO$_{2}$        &   $2.9\tento{-4}$ &   $2.9\tento{-4}$ &   $3.0\tento{-4}$ &   3   & 1 &    50 &  8 &   96  &   9 &   17  &   4 &  108 &  10 \\
C$_{6}$H$_{6}
$  &   $2.2\tento{-4}$ &   $2.3\tento{-4}$ &   $2.3\tento{-4}$ &   3   & 1 &    50 &  8 &   96  &   9 &   17  &   4 &  108 &  10 \\
\hline
\end{tabular}
}
\label{tableMC}
\end{center}
\end{sidewaystable}
%
\newpage
\begin{table*}[h]
\begin{center}
\caption{Mole fractions of lakes species sensitivity to ground total pressure.}
\begin{tabular}{lllrll}
\hline
Compound          & Molar fraction            & Molar fraction              &        & Molar fraction              &    \\
                 & in liquid  at             & in liquid  at               &        & in liquid  at               &    \\
                 & $P_{\rm Huy} = 1467$ hPa  & $P = P_{\rm Huy}\times 0.9$ &        & $P = P_{\rm Huy}\times 1.1$ &    \\
\hline                                                                                                                  \\
 N$_{2}$         & $4.90\times 10^{-3}$      & $4.07\times 10^{-3}$        & -17\%  & $5.92\times 10^{-3}$        &  +21\% \\
 CH$_{4}$        & $9.69\times 10^{-2}$      & $8.26\times 10^{-2}$        & -15\%  & $1.13\times 10^{-1}$        &  +17\% \\
 Ar              & $5.01\times 10^{-6}$      & $4.29\times 10^{-6}$        & -14\%  & $5.83\times 10^{-6}$        &  +16\% \\
 CO              & $4.21\times 10^{-7}$      & $3.49\times 10^{-7}$        & -17\%  & $5.09\times 10^{-7}$        &  +21\% \\
 C$_{2}$H$_{6}$  & $7.64\times 10^{-1}$      & $7.76\times 10^{-1}$        & +1.6\% & $7.50\times 10^{-1}$        & -1.8\% \\
 C$_{3}$H$_{8}$  & $7.42\times 10^{-2}$      & $7.53\times 10^{-2}$        & +1.5\% & $7.28\times 10^{-2}$        & -1.9\% \\
 C$_{4}$H$_{8}$  & $1.39\times 10^{-2}$      & $1.41\times 10^{-2}$        & +1.4\% & $1.37\times 10^{-2}$        & -1.4\% \\
 HCN             & $2.09\times 10^{-2}$ (s)  & $2.27\times 10^{-2}$ (s)    & +8.6\% & $1.91\times 10^{-2}$ (s)    & -8.6\% \\
 C$_{4}$H$_{10}$ & $1.21\times 10^{-2}$ (ns) & $1.23\times 10^{-2}$ (ns)   & +1.7\% & $1.19\times 10^{-2}$ (ns)   & -1.7\% \\
 C$_{2}$H$_{2}$  & $1.15\times 10^{-2}$ (ns) & $1.16\times 10^{-2}$ (ns)   & +0.9\% & $1.13\times 10^{-2}$ (ns)   & -1.7\% \\
 CH$_{3}$CN      & $9.89\times 10^{-4}$ (ns) & $1.00\times 10^{-3}$ (ns)   & +1.1\% & $9.71\times 10^{-4}$ (ns)   & -1.8\% \\
 CO$_{2}$        & $2.92\times 10^{-4}$ (ns) & $2.97\times 10^{-4}$ (ns)   & +1.7\% & $2.87\times 10^{-4}$ (ns)   & -1.7\% \\
 C$_{6}$H$_{6}$  & $2.25\times 10^{-4}$ (ns) & $2.28\times 10^{-4}$ (ns)   & +1.3\% & $2.21\times 10^{-4}$ (ns)   & -1.8\% \\
\hline
\label{influP}
\end{tabular}
\end{center}
\end{table*} 

\newpage
\newpage
\newpage

\newpage

\end{document}